\documentclass[10pt,a4paper,twocolumn,aps,pra,showpacs]{revtex4-1}
\pdfoutput=1
\usepackage[utf8x]{inputenc}
\usepackage{ucs}
\usepackage{amsmath}
\usepackage{amsfonts}
\usepackage{amssymb}
\usepackage{makeidx}
\usepackage{booktabs}
\usepackage{natbib}

\usepackage{microtype}

\usepackage{graphicx}

\renewcommand{\dag}{^{\dagger}}
\newcommand{\exv}[1]{ \langle #1 \rangle }

\newcommand{\ket}[1]{ \lvert #1 \rangle}

\begin{document}

\begin{abstract}
We analyse the possibility to create two-mode spin squeezed states of two separate spin ensembles by inverting the spins in one ensemble and allowing spin exchange between the ensembles via a near resonant cavity field. We investigate the dynamics of the system using a combination of numerical and analytic calculations, and we obtain squeezing for a wide range of parameters. We also investigate the transfer of the squeezing properties to the cavity field and to an output mode from the cavity. Finally, we investigate how the squeezing is affected by effects of inhomogeneities which would be present in solid state implementations of the spin ensembles.
\end{abstract}

\title{Squeezing of Collective Excitations in Spin Ensembles}
\date{\today}
\author{Christian Kraglund Andersen}
\thanks{E-mail: ctc@phys.au.dk}
\author{Klaus Mølmer}
\affiliation{Department of Physics and Astronomy, Aarhus University}


\maketitle

\section{Introduction}

Ensembles of identical quantum systems allow identification of collective degrees of freedom, which may couple strongly to, e.g., radiation fields and thus operate as efficient interfaces for preparation, manipulation and storage of quantum states of light\cite{RevModPhys.82.1041}. We will in this paper study the prospect of creating squeezed quantum states between two separate ensembles of effective two-level systems which both couple to a single quantized field mode of optical or microwave radiation in a cavity.

The basic idea of our proposal is to initialize the first ensemble with all the two-level systems in the ground state and the other ensemble with all systems in the excited state. The coherent transfer of one excitation from one ensemble to the other, mediated by a cavity photon, conserves the energy and total number of excitations in the system, but if we redefine the labelling of the ground and excited states of the inverted ensemble, the excitation transfer is formally equivalent to a simultaneous pair excitation of the ensembles. For a high degree of polarization, collective ensembles of two-level systems are effectively described as harmonic oscillators, where the number operator counts the number of excited two-level systems, while the position and momentum quadrature operators describe collective observables associated with the coherences in the two-level systems or, equivalently, the difference in occupation of superposition states of the system. In a spin language, the number operator is equivalent to the vertical component of the collective spin, while the quadrature operators measure the horizontal collective spin components. Correlated pairwise excitations of two harmonic oscillator modes are known in the non-degenerate optical parametric oscillator (OPO) in quantum optics, where it leads to squeezing and EPR entanglement of optical fields\cite{gardiner00}, and it is the purpose of this manuscript to investigate the accomplishments of the similar process in ensembles of two-level systems.

\begin{figure}[bh]
\includegraphics[width=\linewidth,clip=true,trim=2cm 9cm 2cm 5cm]{{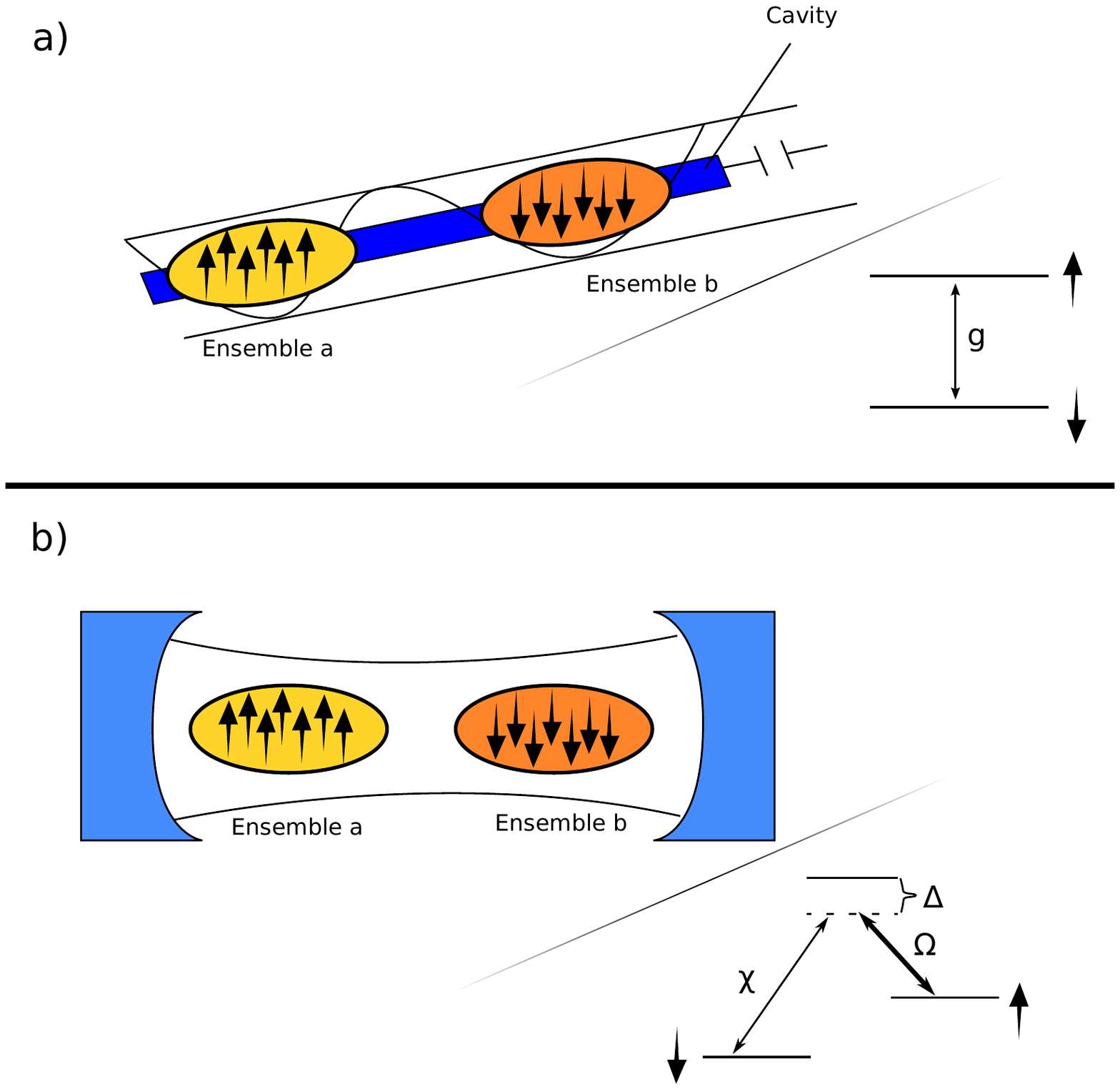}}
\caption{Two physical set-ups for collective spin squeezing of two-level ensembles. a) Two ensembles $a$ and $b$ of two-level systems are prepared in opposite eigenstates, illustrated with spin directions up and down. The spins are situated in a transmission waveguide cavity which can be tuned into resonance with the spin transition. b) Two ensembles of atoms or ions with pairs of ground states and optically excited states are coupled to a classical laser field with Rabi frequency $\Omega$ and to a quantized optical cavity field with coupling strength $\chi$, such that Raman transitions between the ground states are accompanied by the creation or annihilation of a photon in the cavity mode. In both physical implementations the idea is to prepare each ensemble in one of the two internal states, enabling correlated transitions in the two ensembles, by exchange of cavity photons.} \label{system}
\end{figure}

We have two physical systems in mind, as illustrated in Fig.1. In part a) of the figure, we sketch a cavity formed by a transmission waveguide for microwave fields, and we indicate the location of two solid state ensembles with a large number of spins interacting resonantly with the cavity field. Numerous experiments \cite{PhysRevLett.105.140501,PhysRevLett.105.140502,PhysRevLett.105.140503,PhysRevA.85.053806,PhysRevB.84.060501} have demonstrated the strong coupling between a cavity field and spin ensembles, consisting of different dopant ions with both electronic and nuclear spin degrees of freedom, and NV centers in diamond. Following a recent proposal for hybrid quantum computing \cite{PhysRevLett.103.070502}, transfer of quantum states was recently demonstrated between an ensemble of NV centers and a transmon qubit via the cavity field in a similar setup\cite{PhysRevLett.107.220501}. Our present work is related to, and supplements, recent theoretical proposals to generate  squeezed \cite{PhysRevA.85.022324} and entangled states \cite{PhysRevA.85.042306,Lopez:arXiv1111.4966} of NV-center electron spins.

In part b) of the figure, we sketch an optical cavity in which a collection of atoms or ions are trapped and interact with the cavity field and with a classical control field. Two-photon Raman transitions effectively implement atomic ground state changes associated with absorption and emission of single cavity photons. Strong collective coupling to an optical cavity has been observed in recent experiments with ions \cite{Herskind:2009}  and neutral atoms \cite{Colombe2007,Brennecke2007,PhysRevLett.99.213601,citeulike:9736170}. Spin squeezed and entangled states of atomic ensembles have applications in quantum metrology and quantum information protocols \cite{RevModPhys.77.513,duan-2001-414,PhysRevLett.92.013602,FurusawaS1998v282p706}, and protocols for their generation via cavity mediated interactions, different from the one proposed here, have been proposed \cite{PhysRevA.65.053819,PhysRevA.66.022314}.

In this work the ensembles will experience a strong  collectively enhanced coupling to the cavity and, as in  \cite{PhysRevA.85.022324}, the non-classical correlations are established by inverting the population in one of the ensembles such that all the spins are excited and spin flips occur accompanied by excitation of spins in the other ensemble.

In section \ref{sec_Model} we describe the physical systems and the theoretical model that we will use in our calculations. In Sec.  \ref{sec_num} we confirm the squeezing for both short and long times by a combination of numerical calculations and analytical arguments. In Sec. \ref{sec_adi} we consider the limit where the detuning of the cavity is large with respect to the degenerate two-level transition frequencies, and where the cavity can hence be adiabatically eliminated. In Sec.  \ref{sec_out} we show that the squeezing, generated in the collective two-level degrees of freedom, can be released as a squeezed pulse of radiation, and we identify the field mode with the highest degree of squeezing. In Sec. \ref{sec_inhomo}, of particular relevance to solid state spin ensembles, we investigate the robustness of our squeezing and entanglement scheme towards inhomogeneities in the ensembles. Sec.  \ref{sec_conc} concludes the paper.

\begin{figure}[th]
\includegraphics[width=0.92\linewidth,clip=true,trim=1cm 15cm 0.5cm 6cm]{{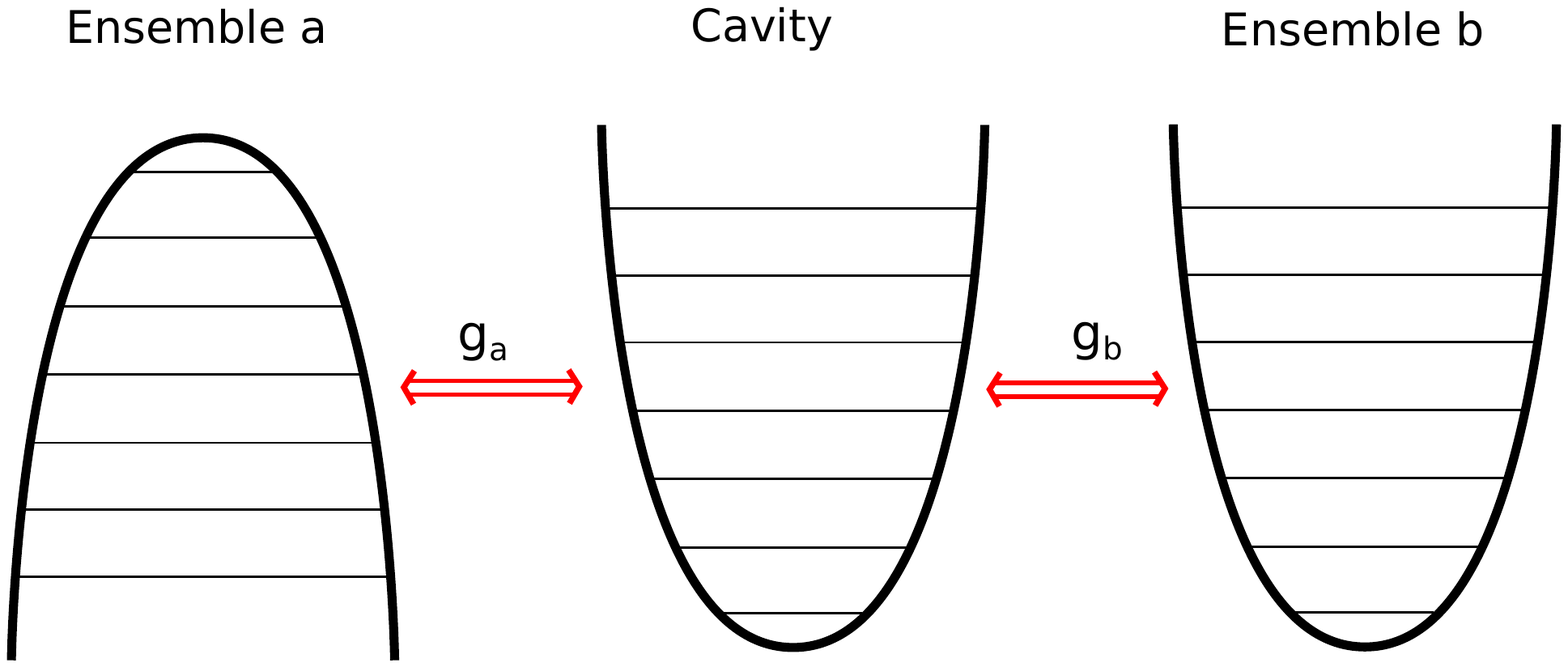}}
\caption{A schematic picture of the system in which we implement spin squeezing. We have three harmonics oscillators of which one is inverted. The cavity oscillator operates as a bus for the transfer of excitations between the two ensemble oscillators.} \label{system2}
\end{figure}

\section{Model of the system} \label{sec_Model}

Assuming no inhomogeneity in the spin system, depicted in Fig. 1a, and choosing $\hbar = 1$, the Hamiltonian for the non-interacting cavity and spin degrees of freedom is given by
\begin{align}
H_0 = \omega_c c\dag c + \frac{1}{2}\sum_{j}^{N_a} \omega_a \sigma_z^{j,a} + \frac{1}{2}\sum_{j}^{N_b} \omega_b \sigma_z^{j,b},
\end{align}
where $\omega_c$ is the angular frequency of the cavity mode with the annihilation (creation) operator $c$($c\dag$), and the Pauli operators, $\sigma_z^{j,n}$, represent the $j$'th spin in ensemble $n$. The cavity mode magnetic field interacts with the spin degree of freedom, and with the rotating wave approximation, the interaction Hamiltonian reads
\begin{align}
H_I = \sum_{j}^{N_a} g_{j,a} (\sigma_+^{j,a} c + \sigma_-^{j,a} c\dag)  + \sum_{j}^{N_a} g_{j,b} (\sigma_+^{j,b} c + \sigma_-^{j,b} c\dag).
\end{align}

In the remaining text we assume equal interaction strengths $g_{j,a} = g_{j,b}=g_1$ for all $j$, and we will apply the Holstein-Primakoff approximation\cite{PhysRev.58.1098} to describe the collective spin degrees of freedom as harmonic oscillators. The $b$-ensemble is prepared with all spins in the spin down ground state, and we define the collective lowering operator $b$ for the ladder of permutation symmetric states of the ensemble with 1, 2, ... flipped spins,
\begin{align}
b = \frac{1}{\sqrt{N_b}} \sum_{j}^{N_b} \sigma_-^{j,b}.
\end{align}
As long as the ensemble has only few excited spins, the ladder of states is well described by the oscillator approximation, $[b,b^\dagger] \simeq 1$,  and the dynamics under the ensuing interaction with the cavity field is readily solvable. This reveals the collectively enhanced coupling due to the large number $N_b$ of spins in the ensemble.

The Holstein-Primakoff approximation applies for small permutation symmetric deviations from any collectively populated state\cite{PhysRevA.81.032314}, and for the inverted $a$-ensemble, with all the spin prepared in the spins up excited state, we define the oscillator lowering (annihilation) operator
\begin{align}
a = \frac{1}{\sqrt{N_a}} \sum_{j}^{N_a} \sigma_+^{j,a}.
\end{align}
The oscillator excitation number then counts the (small) number of spins flipped towards the spin down state, accompanied by the creation of photons in the cavity mode. 

In terms of the oscillator ladder operators, our Hamiltonian of the uncoupled systems writes
\begin{align}
H_0 = \omega_c c\dag c - \omega_a a\dag a + \omega_b b\dag b \label{h_0}
\end{align}
and the interaction Hamilton is 
\begin{align}
H_I = g_a (a\dag c\dag + ca) + g_b (b\dag c + c\dag b) \label{h_i}
\end{align}
with $g_a = \sqrt{N_a} g_1$ and $g_b = \sqrt{N_b} g_1$, revealing the collectively enhanced coupling due to the large numbers, $N_a,$ and $N_b$, of spins in the ensembles. This Hamiltonian can be pictured as shown in Fig. \ref{system2}.

In the case of trapped three-level atoms or ions in an optical cavity, Fig. 1b, we assume an off-resonant classical laser field driving of the transition between the upper atomic ground state, labelled "spin up", and the optically excited state, which in turn is coupled to the lower, "spin down" ground state via the emission of a cavity photon. With a sufficiently large detuning $\Delta$, we may eliminate the optically excited state and thus retrieve a two-photon Raman process between the spin up and down states accompanied by the emission or absorption of cavity photons with a coupling strength $g_1=\Omega\chi/\Delta$, where $\Omega$ is the classical Rabi frequency, and $\chi$ is the coupling strength (per photon) to the quantum field. Ensemble $a$, prepared initially in the spin up state, can undergo spin flip transitions and create cavity photons, while spin flips in ensemble b, prepared in the spin down state are accompanied by absorption of a cavity photon, and this system is thus also described by equation \eqref{h_0} and \eqref{h_i}.

The two set-ups in Fig. 1 involve different physical systems and fields at very different frequencies, and they offer different means of experimental control. The spin ensembles and the microwave cavity may be tuned in and out of resonance with each other thus turning their effective coupling on and off, while the atomic ensembles are driven by a laser field, for which both the intensity and frequency offer means to control the coupling. Cavities for microwave and optical fields typically have different damping times, and while both spins and ground state atoms may have very long lifetimes, solid state spins may experience significant inhomogeneous broadening. We will include cavity damping throughout the following calculations, and we will return to the effects of inhomogeneous broadening in Sec. VII.

The Holstein-Primakoff approximation that we use is only valid as long as the excitation number in each ensemble is small. This will of course have to be checked in the calculations. The magnitude of squeezing can be qualitatively related to the number of quanta involved by recalling the dimensionless position momentum uncertainty relation $\langle x^2 \rangle \langle p^2 \rangle \geq 1/4$ and the energy relation $\langle p^2/2 + x^2/2\rangle = \langle n \rangle +\frac{1}{2}$, which suggests an amplitude squeezing factor $\sim 1/\langle n \rangle$. Spin squeezing by a factor 10-50 thus involves flipping of a similar number of spins, which is indeed much less than the number of spins that we have in mind for the ensembles.

The interaction Hamiltonian includes a term resembling the usual non-degenerate OPO Hamiltonian\cite{gardiner00}, and in the absence of the $b$-ensemble, we would expect the rather straightforward formation of a two-mode squeezed state of the cavity field and the $a$-ensemble. Since the cavity field leaks, this state will not live for long, and we hence couple the field to the second $b$-ensemble and investigate if a long lived squeezed state can be created between the two ensembles.

\begin{figure}[t]
\includegraphics[width=\linewidth,clip=true,trim=2cm 17.6cm 3cm 1.5cm]{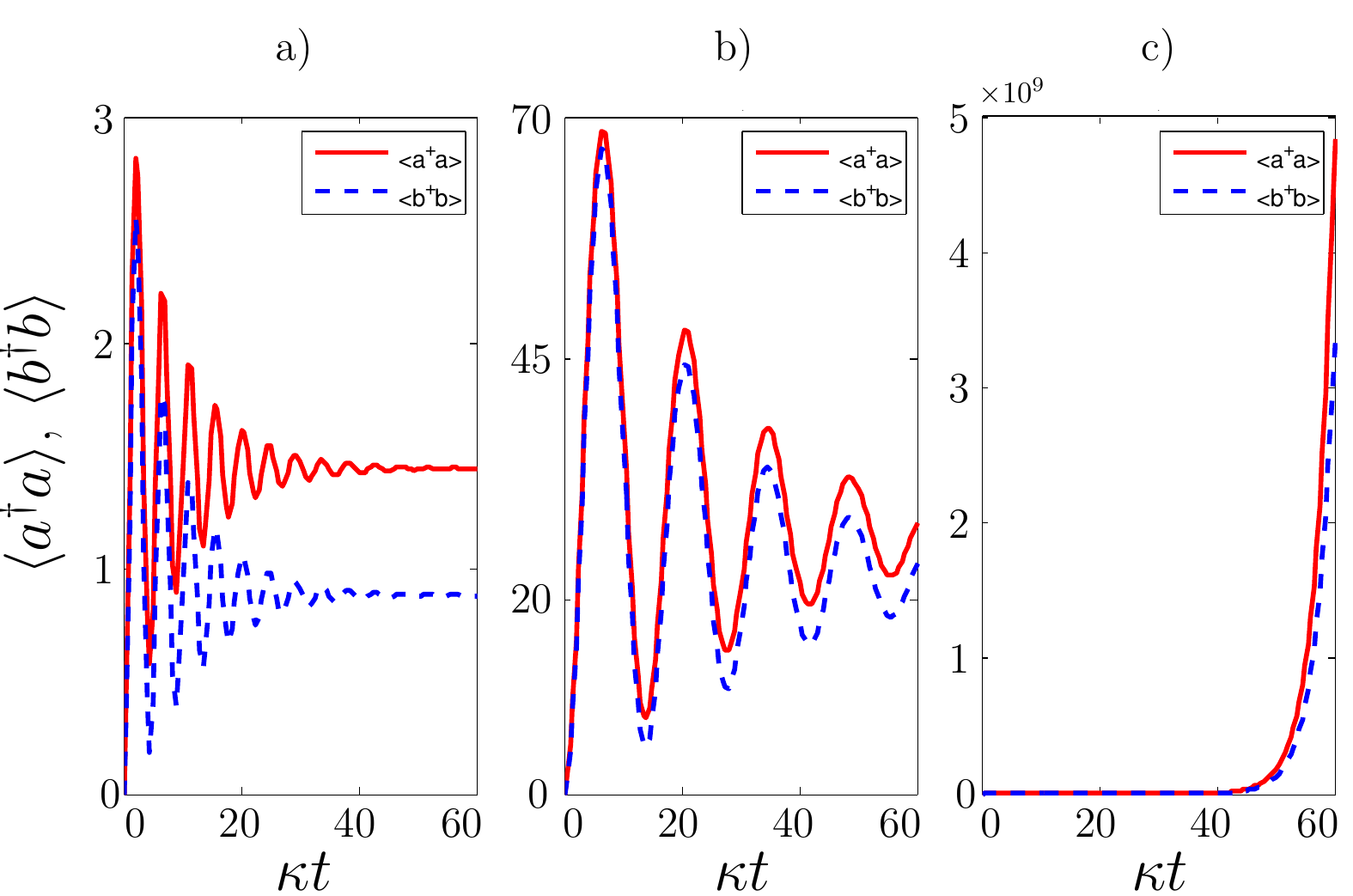}
\caption{The average number of excitations in the two ensembles calculated numerically with the parameters  $\Delta = \omega' - \omega = 10\kappa$, $ g_b = 5\kappa$, and with $g_a = 0.6g_b$, $g_a = 0.9 g_b$, and  $g_a =  1.2 g_b$, in panels a, b and c, respectively.} \label{ensemble_twocases}
\end{figure}

We first assume that $g_a,g_b > \kappa$, where $\kappa$ is the field decay rate. In this case the dynamics between the oscillators is much faster than the damping of the cavity and for short times we can therefore neglect the decay term.
For a simple and qualitative argument lets consider the resonant case, $\omega_c = \omega_a = \omega_b$, where the time evolution operator in the interaction picture is given by
\begin{align}
\mathcal{U}(dt) = e^{-iH_I dt} \approx \mathbb{I} - iH_I dt - \frac{1}{2} H_I^2 dt^2. \label{u_dt}
\end{align}
Applying \eqref{u_dt} to the initial state, we get
\begin{align}
&\mathcal{U}(dt) \ket{0,0,0} =\ket{0,0,0} - i g_a dt \ket{1,0,1} \nonumber \\ &- \frac{dt^2}{2} (2 g_a^2 \ket{2,0, 2} + g_a^2 \ket{0,0,0} + g_a g_b \ket{1,1,0} ), \label{sq_state}
\end{align}
and we observe the last component with a simultaneous excitation in both ensemble oscillators, characteristic for two-mode squeezing. The variance of the two-mode quadrature operator 
\begin{align}
X_{ab}(\theta) = \frac{1}{2} \Big(\frac{e^{i\theta}}{\sqrt{2}} (a + b) + \frac{e^{-i\theta}}{\sqrt{2}} (a\dag + b\dag ) \Big),
\end{align}
can be readily obtained, and for $\theta = 0$ we get
\begin{align}
\exv{(\Delta X_{ab}(0))^2} = \frac{1}{4} + \frac{dt^2}{2}(3g_a^2 - 4g_ag_b). \label{shortXsq}
\end{align}
We clearly see a small reduction, provided $g_b > 3g_a/4$, where the asymmetry in the requirements on the coupling strengths is due to the respective OPO and beam-splitter like couplings to the ensembles.

\begin{figure}[t]
\includegraphics[width=\linewidth,clip=true,trim=2cm 19.7cm 1.4cm 1.5cm]{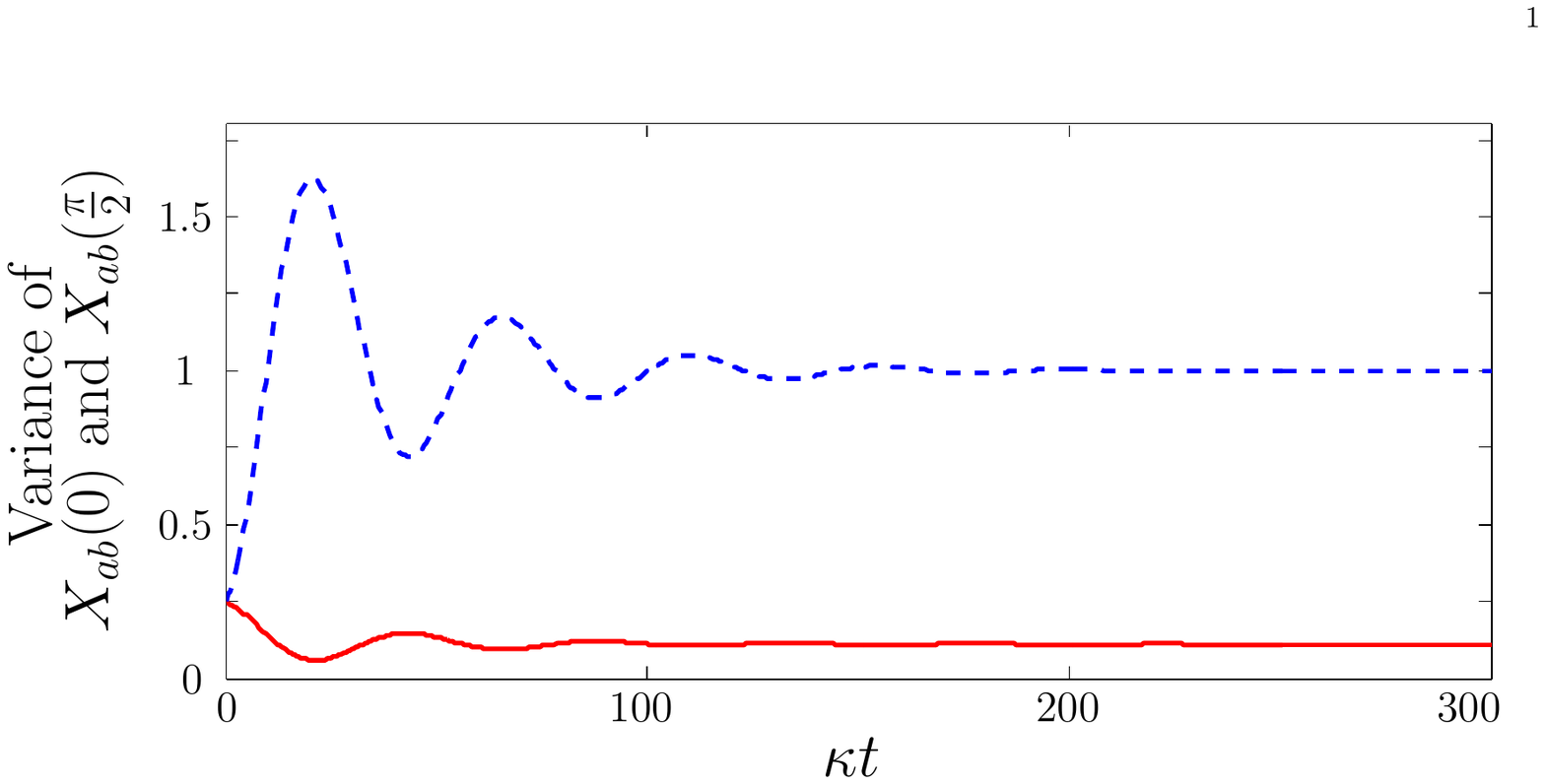}
\caption{Variances of the collective quadrature observables $X_{ab} (0)$ (red solid line), and $X_{ab}(\pi /2)$ (blue dashed line) for the coupling parameters and detuning, $g_b = \kappa$ , $\Delta = 5 \kappa$ and $g_a=0.5g_b$.} \label{ensemble_squeezing}
\end{figure}

\begin{figure}[t]
\includegraphics[width=\linewidth,clip=true,trim=2cm 19.7cm 1.6cm 1.5cm]{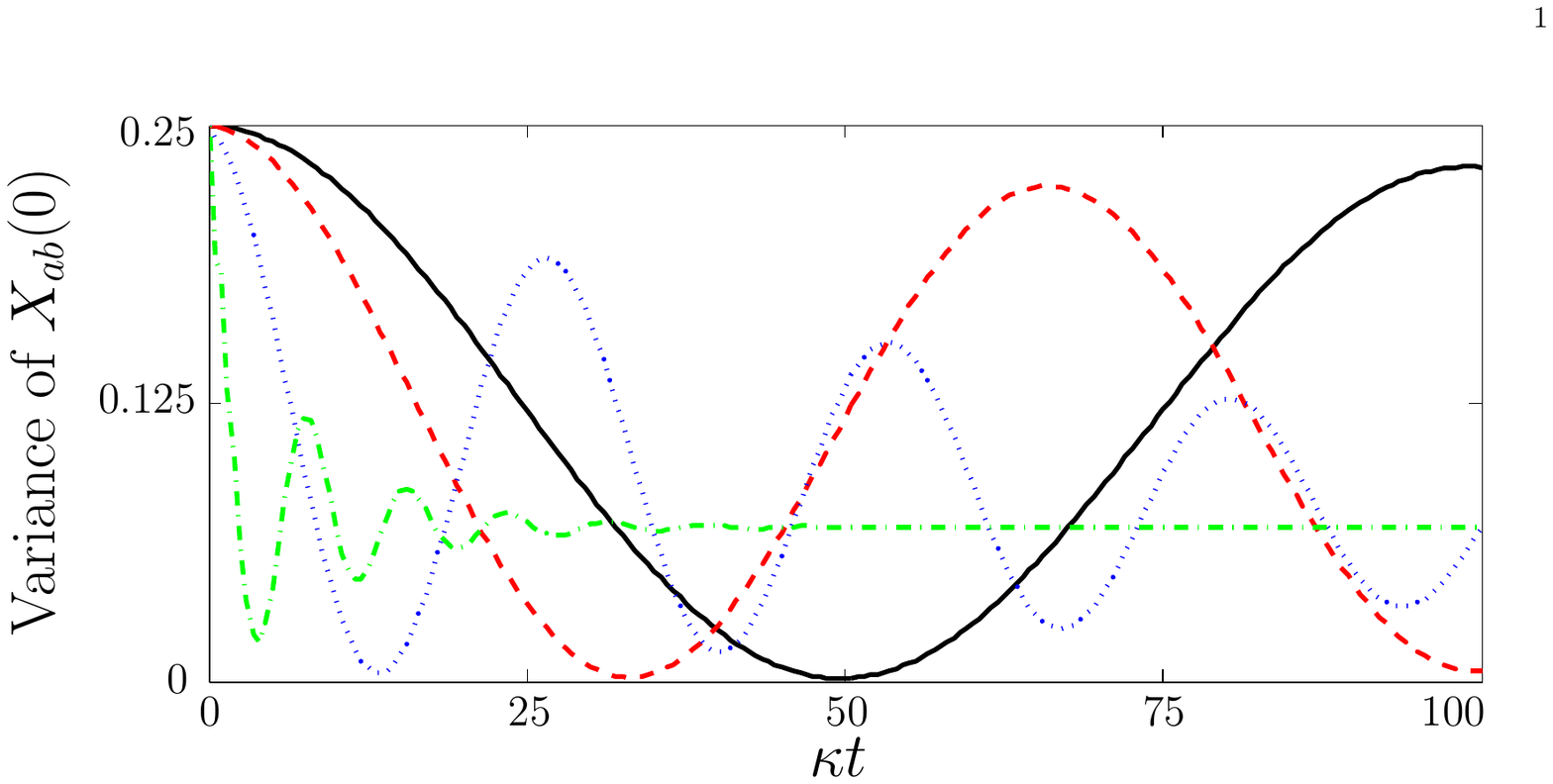}
\caption{The variance of $X_{ab} (0)$ for different detuning of the cavity. We have $g_b = 5 \kappa$ and $g_a = 0.9g_b$ . The black (solid) line is for $\Delta = 75\kappa$. The red (dashed) line is for $\Delta = 50\kappa$. The blue (dotted) line is $\Delta = 20\kappa$ and the green (dashed dotted) line is $\Delta = 5\kappa$.} \label{detuned_squeezing}
\end{figure}

\section{Heisenberg-Langevin equations, damping and steady state} \label{sec_num}

The bilinear Hamiltonian leads to simple Heisenberg equations of motion for the oscillator ladder operators. Cavity damping is modelled as a similar coupling to the continuum of free-space vacuum modes at one of the cavity mirrors, leading to a damping rate $\kappa$ of the intra-cavity field accompanied by a white-noise term \cite{gardiner00},
\begin{subequations}
\begin{align}
\frac{d a}{dt} &= i\omega a - ig_ac\dag \\
\frac{d a\dag}{dt} &= -i \omega a\dag + ig_ac \\
\frac{d b}{dt} &= -i\omega b - ig_bc \\
\frac{d b\dag}{dt} &= i\omega b\dag + ig_b c\dag \\
\frac{d c}{dt} &= -(\kappa + i \omega' ) c - i g_a a\dag -ig_b b - \sqrt{2\kappa} c_{in}(t) \\
\frac{d c\dag}{dt} &= -(\kappa - i \omega') c\dag + ig_a a + i g_b b\dag + \sqrt{2\kappa} c_{in}\dag(t),
\end{align} \label{langevin_homo}
\end{subequations}
with $[ c_{in}(t), c_{in}\dag(t') ] = \delta (t-t')$. In these equations we assume that the spin ensembles have degenerate excitation frequencies $\omega$ and the cavity field may have a different frequency $\omega'$. Starting from the oscillator ground states, all amplitudes have zero mean for all times, and the state is fully characterized by the second moments of the quadrature operators, which solve a linear coupled set of equations, following in a straightforward manner from \eqref{langevin_homo}. These equations are readily solved numerically, and in Fig. \ref{ensemble_twocases} we show the average excitation number of the two ensemble oscillators for different parameters. The mean excitation of the ensembles show two different regimes: One in which the excitation number undergoes a damped oscillation and one in which it increases exponentially. 

We can understand this behaviour if we define two new superposition mode annihilation operators
\begin{align}
d = (g_b a + g_a b\dag)/\sqrt{g_b^2-g_a^2} \label{homo_dark} \\
e = (g_a a\dag + g_b b)/\sqrt{g_b^2-g_a^2},
\end{align}
assuming $g_b > g_a$. It follows from \eqref{langevin_homo} that  
\begin{align}
\frac{d \, }{d t} d= i\omega d,
\end{align}
and this "dark" mode $d$ is uncoupled from the dynamics with a trivial time evolution. If we go to a rotating frame at the ensemble frequency $\omega$, $\frac{d }{d t} d = 0$, then we can easily find the eigenvalues of the matrix that represents the homogeneous part of equation \eqref{langevin_homo},
\begin{subequations}
\begin{align}
\lambda_1 &= 0 \\
\lambda_2 &= 0 \\
\lambda_3 &= \frac{i\Delta}{2} - \frac{\kappa}{2} - \frac{1}{2}\sqrt{G^2} \\
\lambda_4 &= \frac{i\Delta}{2} - \frac{\kappa}{2} + \frac{1}{2}\sqrt{G^2} \\
\lambda_5 &= -\frac{i\Delta}{2} - \frac{\kappa}{2} - \frac{1}{2}\sqrt{G^2} \\
\lambda_6 &= -\frac{i\Delta}{2} - \frac{\kappa}{2} + \frac{1}{2}\sqrt{G^2}
\end{align} \label{homo_eig}
\end{subequations}
with $G^2 = 4g_a^2- 4g_b^2  + \kappa^2 - \Delta^2 - 2i\kappa\Delta$, where $\Delta = \omega' - \omega$. The homogeneous solutions can be written as linear combinations of $e^{\lambda_i t}$, and for $\Delta = 0$ and $g_b^2 > g_a^2 + \kappa^2$, the real part of $\sqrt{G^2}$ is smaller than $\kappa$, ensuring a damped oscillatory behaviour of the solutions. 

\begin{figure}[t]
\includegraphics[width=\linewidth,clip=true,trim=2.0cm 17.7cm 2.0cm 1.5cm]{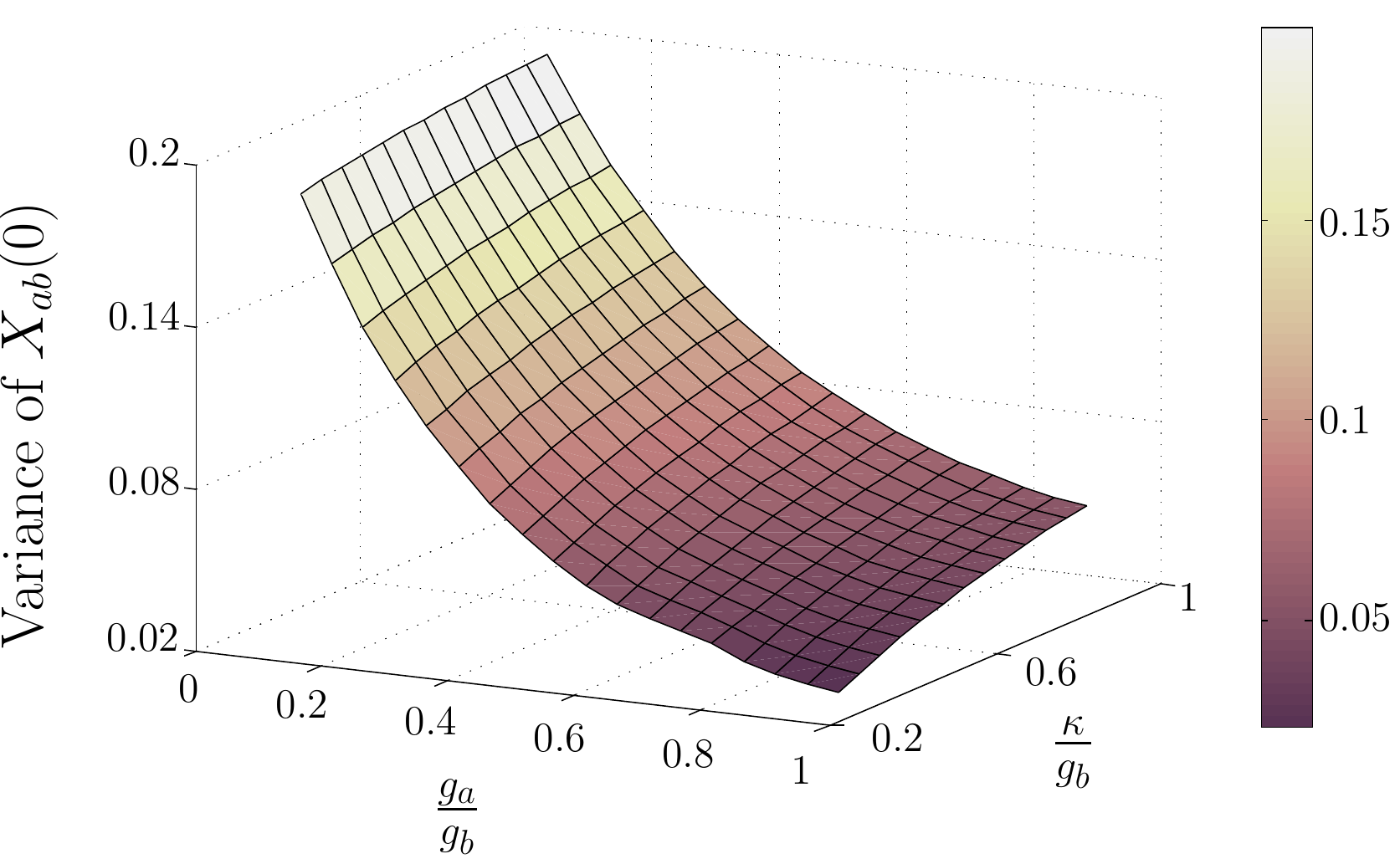}
\caption{The minimally achieved variance for $\Delta = g_b$ and different values of $\kappa$ and $g_a$.} \label{squeezing_minimum}
\end{figure}

In the parameter regime of a damped oscillatory solution, our system thus reaches a steady state, in which we will now investigate the squeezing properties. In Fig. \ref{ensemble_squeezing} we show the variance of $X_{ab}(0)$ and $X_{ab}(\frac{\pi}{2})$, and we see that in steady state we obtain a squeezed state. The amount of squeezing depends on the detuning, the coupling strength and the decay rate of the cavity. In Fig. \ref{detuned_squeezing} we show the variance of $X_{ab}(0)$ for different values of the detuning and we observe squeezing for both short and large times. For large cavity detuning the damping due to cavity loss is small, and the spin ensembles show an oscillatory dynamic between a significantly squeezed and a non-squeezed state, while for smaller detuning the damping is more pronounced, and we quickly reach a squeezed steady state. In Fig. \ref{squeezing_minimum} we show the minimally achieved variance for different values of the coupling strengths to the inverted ensemble and cavity decay rate. Small damping and strong coupling lead to the largest degree of collective squeezing.

\section{Large detuning - adiabatic elimination of cavity field} \label{sec_adi}

\begin{figure}[tb]
\includegraphics[width=\linewidth,clip=true,trim=2cm 19.5cm 1.6cm 1.5cm]{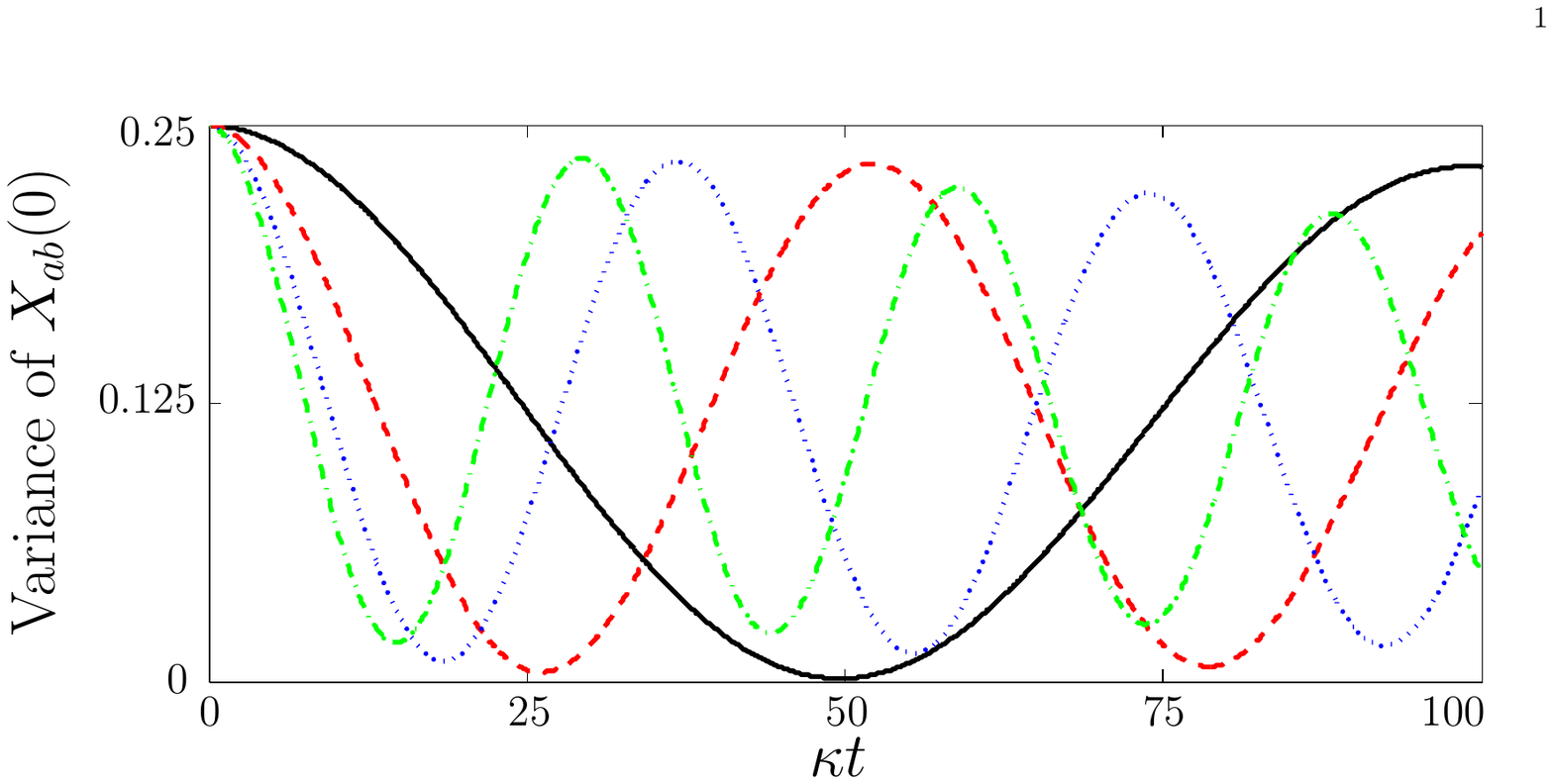}
\caption{The collective quadrature variance, determined by adiabatic elimination of the detuned cavity field, \eqref{variance_analytic}, with $\Delta = 75\kappa$ and $g_b = 5\kappa$. We vary the coupling strength, while remaining in the adiabatic regime. For the black (solid) line we choose $g_a = 0.9g_b$, corresponding to the black line in Fig. \ref{detuned_squeezing}. For the red (dashed) line we choose $g_a=0.8g_b$, the blue (dotted) line $g_a=0.7g_b$ and the green (dashed dotted) $g_a = 0.6 g_b$.} \label{adiabatic_squeezing}
\end{figure}

If the cavity field detuning is large, $\Delta = \omega' - \omega \gg g_a, g_b$, we may eliminate the field operators from the problem by setting the derivative to zero
\begin{align}
0 = \frac{d c}{dt} = -(\kappa + i \Delta ) c - i g_a a\dag -ig_b b - \sqrt{2\kappa} c_{in}(t)
\end{align}
with the solution
\begin{align}
c(t) = - i \frac{g_a a\dag + g_b b}{\kappa + i\Delta} + \frac{\sqrt{2\kappa}}{\kappa+i\Delta} c_{in}(t).
\end{align}
This expression can be inserted into the Heisenberg-Langevin equations for $a$ and $b$ such that we get
\begin{subequations}
\begin{align}
\frac{d a}{dt} &= \frac{g_a^2}{\kappa-i\Delta} a + \frac{g_a g_b}{\kappa-i\Delta} b\dag - \frac{ig_a\sqrt{2\kappa}}{\kappa-i\Delta} c_{in}\dag(t) \\
\frac{d a\dag}{dt} &= \frac{g_a^2}{\kappa+i\Delta} a\dag + \frac{g_a g_b}{\kappa+i\Delta} b + \frac{ig_a\sqrt{2\kappa}}{\kappa+i\Delta} c_{in}(t) \\
\frac{d b}{dt} &= -\frac{g_b^2}{\kappa+i\Delta} b - \frac{g_a g_b}{\kappa+i\Delta} a\dag - \frac{ig_b\sqrt{2\kappa}}{\kappa+i\Delta} c_{in}(t) \\
\frac{d b\dag}{dt} &= -\frac{g_b^2}{\kappa-i\Delta} b\dag - \frac{g_a g_b}{\kappa-i\Delta} a + \frac{ig_b\sqrt{2\kappa}}{\kappa-i\Delta} c_{in}\dag(t).
\end{align} \label{langevin_adiabatic} \end{subequations}
The reduced dimensionality of the problem now allows a straightforward solution, and, e.g., for $a(t)$ we have
\begin{align}
a(t) &= \frac{g_a^2 e^{t\frac{g_a^2 - g_b^2}{\kappa - i \Delta}} - g_b^2}{g_a^2 - g_b^2} a(0) + \frac{g_ag_b e^{t\frac{g_a^2 - g_b^2}{\kappa - i \Delta}} - g_bg_b}{g_a^2 - g_b^2} b\dag(0) \nonumber \\
&+ \frac{i\sqrt{2\kappa}}{\kappa-i\Delta} \int_0^t \Bigg(g_a\frac{g_a^2 e^{(t-t')\frac{g_a^2 - g_b^2}{\kappa - i \Delta}}- g_b^2}{g_a^2 - g_b^2} c_{in}\dag(t') \nonumber \\
& - g_b \frac{g_ag_b e^{(t-t')\frac{g_a^2 - g_b^2}{\kappa - i \Delta}} - g_ag_b}{g_a^2 - g_b^2} c_{in}\dag(t') \Bigg) dt'.
\end{align}
Similar expressions can be found for $a\dag$, $b$ and $b\dag$, and we can thus compute the variance of the quadrature $ X_{ab}(0)$ observable, 
\begin{align}
\exv{(\Delta X_{ab}(0))^2} = \frac{1}{8} \Bigg(& \frac{g_a^2 + g_b^2}{(g_a + g_b)^2} \left(1 + e^{2 \kappa t \frac{g_a^2 - g_b^2}{\kappa^2 + \Delta^2} } \right) \nonumber \\
&+ \frac{2g_ag_b \big(e^{t\frac{g_a^2 - g_b^2}{\kappa + i \Delta}} + e^{-t\frac{g_a^2 - g_b^2}{i \Delta - \kappa }}\big) }{(g_a+g_b)^2} \Bigg) \nonumber \\
& + f(t), \label{variance_analytic}
\end{align}
where $f(t)$ is the contribution from the input field noise-terms. This is a double integral over the noise operators $c_{in}$ and $c_{in}\dag$, which by the commutator relations reduces to a single integral, which evaluates to
\begin{align}
 f(t) = \frac{\left(e^{2 \kappa t \frac{(g_a^2-g_b^2) }{\kappa^2 + \Delta^2} } - 1\right) \left(g_a - g_b\right) }{8\left(g_a + g_b\right)}.
 \end{align}
In Fig. \ref{adiabatic_squeezing} we have plotted equation \eqref{variance_analytic} for different coupling strength and wee see that the behaviour resembles the result from the numeric analysis.

The exponential terms in equation \eqref{variance_analytic} and in the integral over the noise terms are decaying with time, if $g_b > g_a$, and in the long time limit the variance converges to
\begin{align}
\exv{(\Delta X_{ab}(0))^2}_{t\rightarrow \infty} =  \frac{g_b^2}{4(g_a + g_b)^2}, \label{variance_long}
\end{align}
which also confirms the squeezing results. 

Furthermore we know that the exponents in \eqref{variance_analytic} are small in the adiabatic regime, which enables an expansion of the exponential functions such that we get an approximate expression for the variance,
\begin{align}
\exv{(\Delta X_{ab}(0))^2} \approx  \frac{1}{4} \Bigg(1 + \frac{g_a^2 - g_b^2}{\kappa^2 + \Delta^2} \kappa t \Bigg),
\end{align}
which is valid for small times only.

If we choose a finite interaction time $t$, such that $t \frac{g_b^2 - g_a^2}{\Delta} = \pm \pi$, and if $\Delta \gg \kappa$, the complex exponential terms in \eqref{variance_analytic} undergo a complete change of sign before any appreciable damping of the exponential functions occurs. To first order in $\frac{\kappa}{\Delta}$ this leads to a minimum in the variance 
\begin{align}
\exv{(\Delta X_{ab}(0))^2}_{min} =  \frac{(g_a - g_b)^2}{4(g_a + g_b)^2} +  \Bigg\lvert \frac{2g_a(g_b - g_a)}{8(g_a+g_b)^2} \frac{\kappa}{\Delta} \Bigg\rvert, \label{variance_min}
\end{align}
which approaches zero for $g_a$ approaching $g_b$ and $\frac{\kappa}{\Delta}$ approaching zero. These minima are readily observed in \mbox{Fig. \ref{adiabatic_squeezing}}.

\section{Squeezing in the Output Field} \label{sec_out}

Conventional squeezed light sources are based on optically non-linear materials and they are subject to a competition between the non-linear processes and losses. In our scheme, the squeezing of the spin ensembles is not due to a non-linear interaction but to an exchange of quanta with an initially inverted medium, and our analysis suggests that a  potentially high degree of squeezing is obtainable. This squeezing may be relevant for entanglement operations on the spin degrees of freedom, but they may also be coupled resonantly to the cavity field, and thus lead to production of a squeezed output field from the cavity. To accomplish this mapping, we will apply a $\pi$-pulse to the inverted ensemble, such that its excitations, correlated with the ones in the other ensemble, can be converted into photons according to the Hamiltonian of the system, where the annihilation and creation operators of ensemble $a$ now attain the same roles as in ensemble $b$,  
\begin{align}
H'_0 &= \omega_c c\dag c + \omega_a a\dag a + \omega_b b\dag b \\
H'_I &= g_a (a\dag c + c\dag a) + g_b (b\dag c + c\dag b). \label{h_i2}
\end{align}
We observe that the spin ensembles also here display an uncoupled collective mode, while the collective mode with annihilation operator $\propto g_a a + g_b b$ couples directly to the cavity field. Ideally, the squeezing properties in this mode should be converted into cavity photons leaving the cavity at a rate $\kappa$ and thus forming a finite pulse with favourable squeezing properties.

The Hamiltonian coupling and the damping and input noise fields leads to  Heisenberg-Langevin equations for cavity field and the coupled mode, that we can write on the form 
\begin{align}
\frac{d}{dt} \mathbf{r} (t) &= M_{in}\mathbf{r}(t) -\sqrt{2\kappa} \mathbf{r}_{in}(t), 
\end{align}
where
\begin{align}
\mathbf{r}(t) = \begin{pmatrix}
g_a a(t) + g_b b(t) \\ c(t)
\end{pmatrix}
\end{align}
and 
\begin{align}
\mathbf{r}_{in}(t) = \begin{pmatrix} 0 \\ c_{in}(t)\end{pmatrix},
\end{align}
and the matrix $M_{in}$ contains the coupling terms from the Heisenberg-Langevin equations.

The input/output formalism \cite{gardiner00, PhysRevA.30.1386}, relates the input and output fields by the equation
\begin{align*}
c_{out}(t) - c_{in}(t) = \sqrt{2\kappa} c(t)
\end{align*}
This allows us to write the Heisenberg-Langevin equations in an alternative form, driven by the output noise operators,
\begin{align}
\frac{d}{dt} \mathbf{r}(t) &= M_{out}\mathbf{r}(t) -\sqrt{2\kappa} \mathbf{r}_{out}(t),
\end{align}
where
\begin{align}
\mathbf{r}_{out}(t) = \begin{pmatrix} 0 \\ c_{out}(t)\end{pmatrix},
\end{align}
and $M_{out}$ differs from $M_{in}$ by a change of sign of the damping term of $c$.

The Laplace-transform turns the differential equations into algebraic forms
\begin{align}
s \tilde{\mathbf{r}}(s) - \mathbf{r}(0) &= M_{in} \tilde{\mathbf{r}}(s) - \sqrt{2\kappa} \tilde{\mathbf{r}}_{in}(s) \\
&=  M_{out} \tilde{\mathbf{r}}(s) - \sqrt{2\kappa} \tilde{\mathbf{r}}_{out}(s),
\end{align}
with $t=0$ at the time of the $\pi$-pulse. From this we can eliminate $\tilde{\mathbf{r}}(s)$ and obtain the expression
\begin{align}
\tilde{\mathbf{r}}_{out}(s) =& (s-M_{out})(s-M_{in})^{-1}\frac{ \big(\sqrt{2\kappa} \tilde{\mathbf{r}}_{in}(s) - \mathbf{r}(0) \big) }{\sqrt{2\kappa}} \nonumber \\
& + \frac{\mathbf{r}(0)}{\sqrt{2\kappa}}.
\end{align}
By carrying out the matrix inversion and mulitplications, and performing the inverse Laplace-transform we finally obtain the expression
\begin{align}
c_{out} (t) =& \frac{\alpha(t)(g_a a(0) + g_b b(0))- \beta(t) c(0)}{\sqrt{2\kappa}}  \nonumber \\&+ \int_0^{t} \beta(t-t') c_{in}(t') \, dt' + c_{in}(t),
\end{align}
with the functions
\begin{align}
\alpha(t) &= \frac{4i\kappa}{\Omega} e^{-\frac{1}{2}(\kappa+i\Delta) t} \sinh{\big(\frac{1}{2} \Omega t\big)} \\
\beta(t) &= 2\kappa\Big(\frac{\kappa + i\Delta}{\Omega} \sinh{\big(\frac{1}{2} \Omega t\big)} -\cosh{\big(\frac{1}{2} \Omega t\big)} \Big) e^{-\frac{1}{2}(\kappa+i\Delta) t}
\end{align}
where $\Omega = \sqrt{\kappa^2  -4g_a^2- 4g_b^2  - \Delta^2 - 2i\kappa\Delta}$.
With the expression for the output-field in terms of operators with known noise properties, we can compute the degree of squeezing at any moment of time and non-classical correlations between different times. Since the field is emitted in a pulse after the inversion of the $a$ ensemble, we expect that a time dependent mode function for the field $u(t)$ exists, for which the squeezing is maximal. We define the corresponding mode operator\cite{PhysRevA.73.063804} 
\begin{align}
c_u = \int_0^{\infty} u(t) c_{out}(t) \, dt
\end{align}
where $u(t)$ is normalized such that
\begin{align}
\big[c_u \, , \, c_u\dag \big] = 1.
\end{align}
The noise in the corresponding field quadratures is measured by homodyne
detection with either a local oscillator with the appropriate time
dependence, or it is determined by a weighted temporal integral of the homodyne signal obtained with a constant local oscillator. The degree of squeezing of course depends on the shape of the chosen mode function u(t), and we have studied different candidate mode functions, including a simple exponential decay with rate $\kappa$, the real square root of the mean intra cavity photon number\cite{1205.6004}, and a general complex  function u(t) which is numerically optimized to yield the largest degree of squeezing. 

\begin{figure}[tbh]
\includegraphics[width=\linewidth,clip=true,trim=2cm 18cm 1.1cm 1.5cm]{{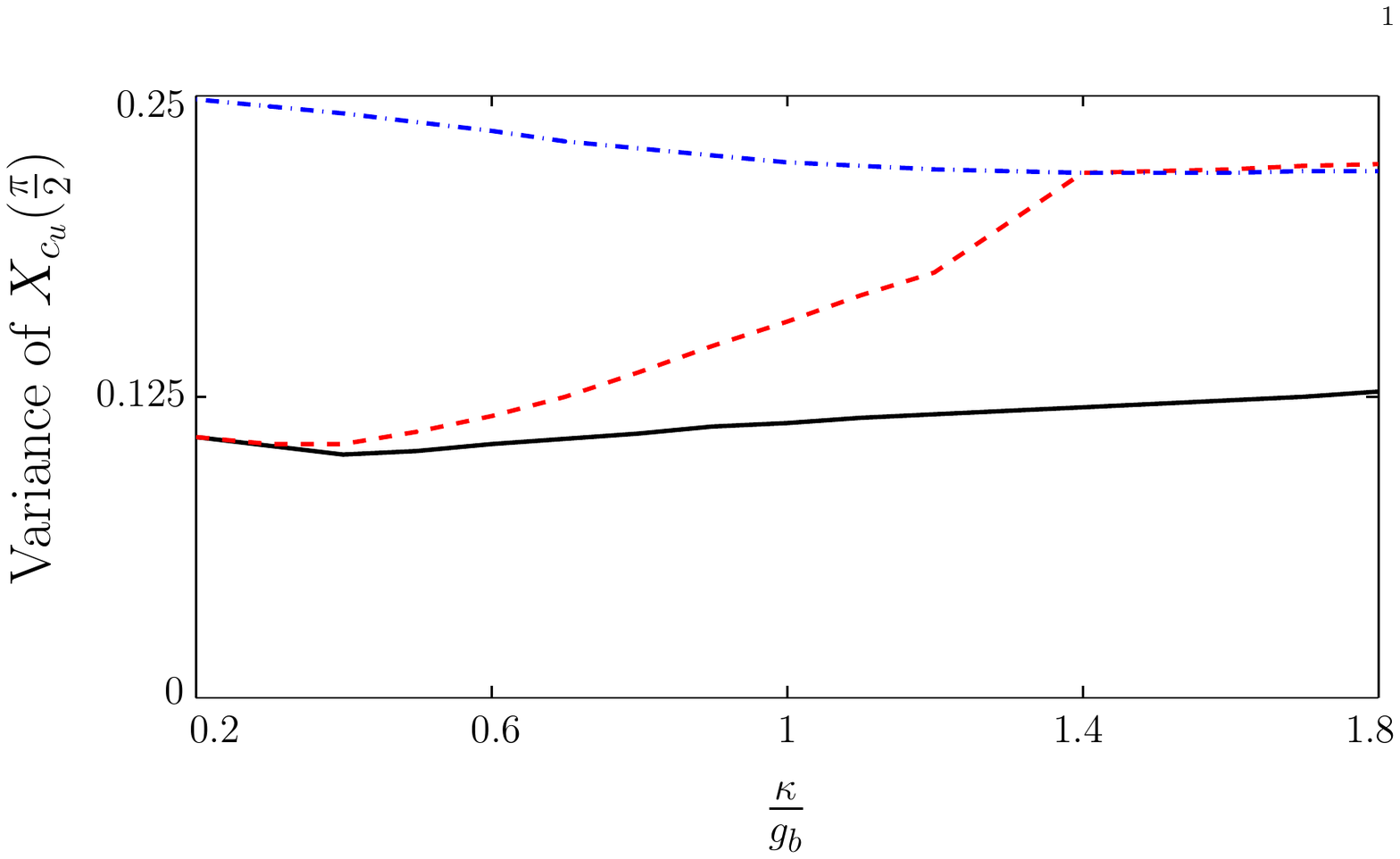}}
\caption{The output field quadrature variance with $u(t) = Ae^{-\kappa t/2}$ (blue dashed dotted), $u(t) = B (\exv{c\dag c})^{1/2}$ (red dashed) and a optimized $u(t)$ (solid black line) for different values of $\kappa$. We take $\Delta_c = 0.01g_b$ and $g_a = 0.9 g_b$.} \label{output_ut_k}
\end{figure}

\begin{figure}[tbh]
\includegraphics[width=\linewidth,clip=true,trim=2cm 18.7cm 1.1cm 2cm]{{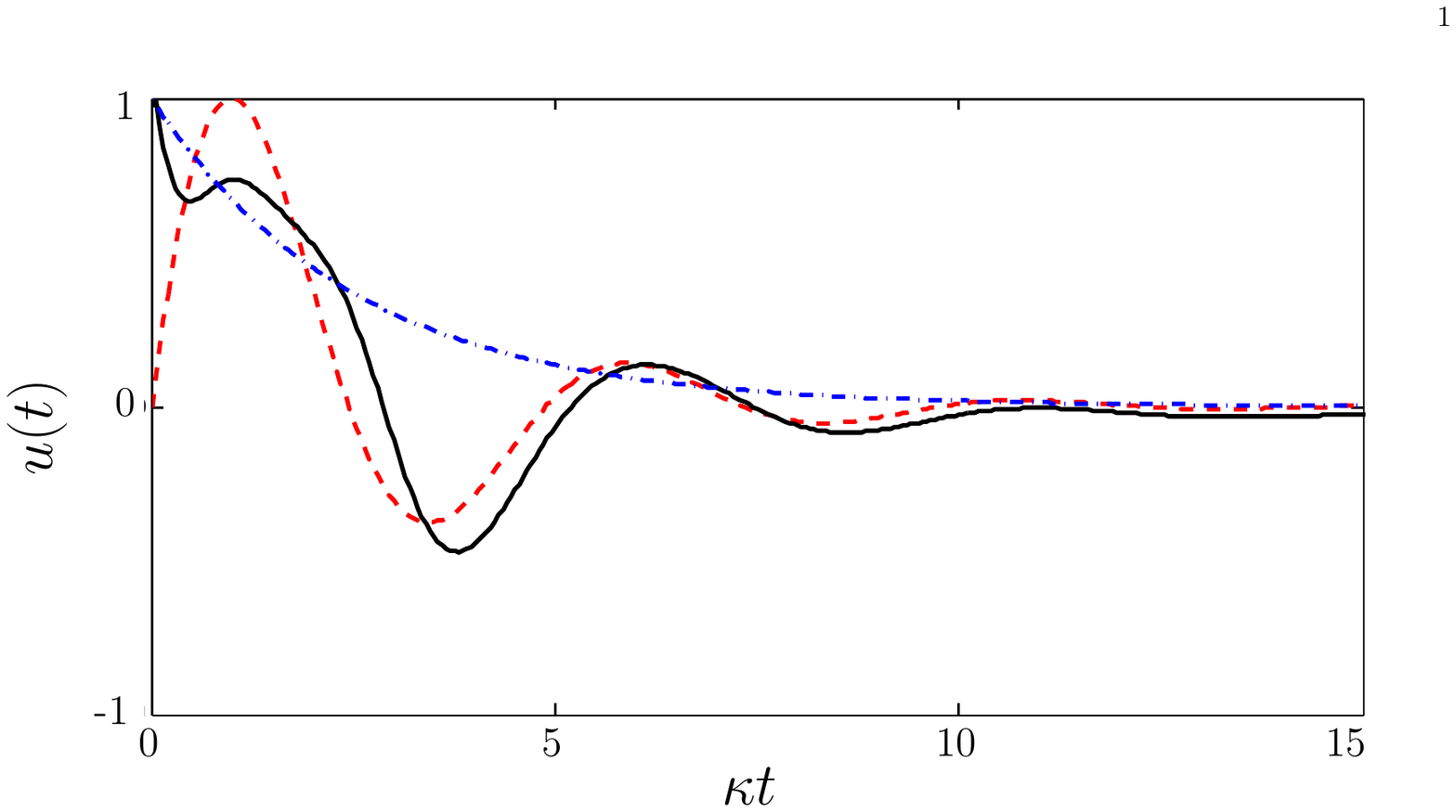}}
\caption{Non-normalized $u(t)$. We have $u(t) = Ae^{-\kappa t/2}$ (blue dashed dotted), $u(t) = A (\exv{c\dag c})^{1/2}$ (red dashed) and optimized $u(t)$ (solid black line). We take $g_b = 1.25 \kappa$, $\Delta = 0.001\kappa$ and  $g_a = 0.9 g_b$.} \label{ut_k08}
\end{figure}

The variance of $X_{c_u}(\frac{\pi}{2})$ for these three choices of mode functions are shown in Fig. \ref{output_ut_k}, and examples of their time dependence are illustrated in Fig. \ref{ut_k08}. We note that if the emitted light is squeezed within a single mode, the choice of a different mode function is equivalent to the admixing with a vacuum state with a standard variance of $\frac{1}{4}$. The exponentially decaying function thus has a too small overlap with the optimum mode to show any significant squeezing. For small damping, the square root of the emitted power provides a good ansatz for the squeezed light mode. For stronger cavity damping the numerically optimized mode deviates in particular for early times and shows significantly better squeezing. The dependence of the optimal squeezing on kappa in Fig. \ref{output_ut_k} is similar to the spin squeezing dependence shown in Fig.  \ref{squeezing_minimum}, and suggests that the squeezing properties are successfully transferred to the emitted light.

\begin{figure}[th]
\includegraphics[width=\linewidth,clip=true,trim=2cm 19.7cm 1.5cm 2cm]{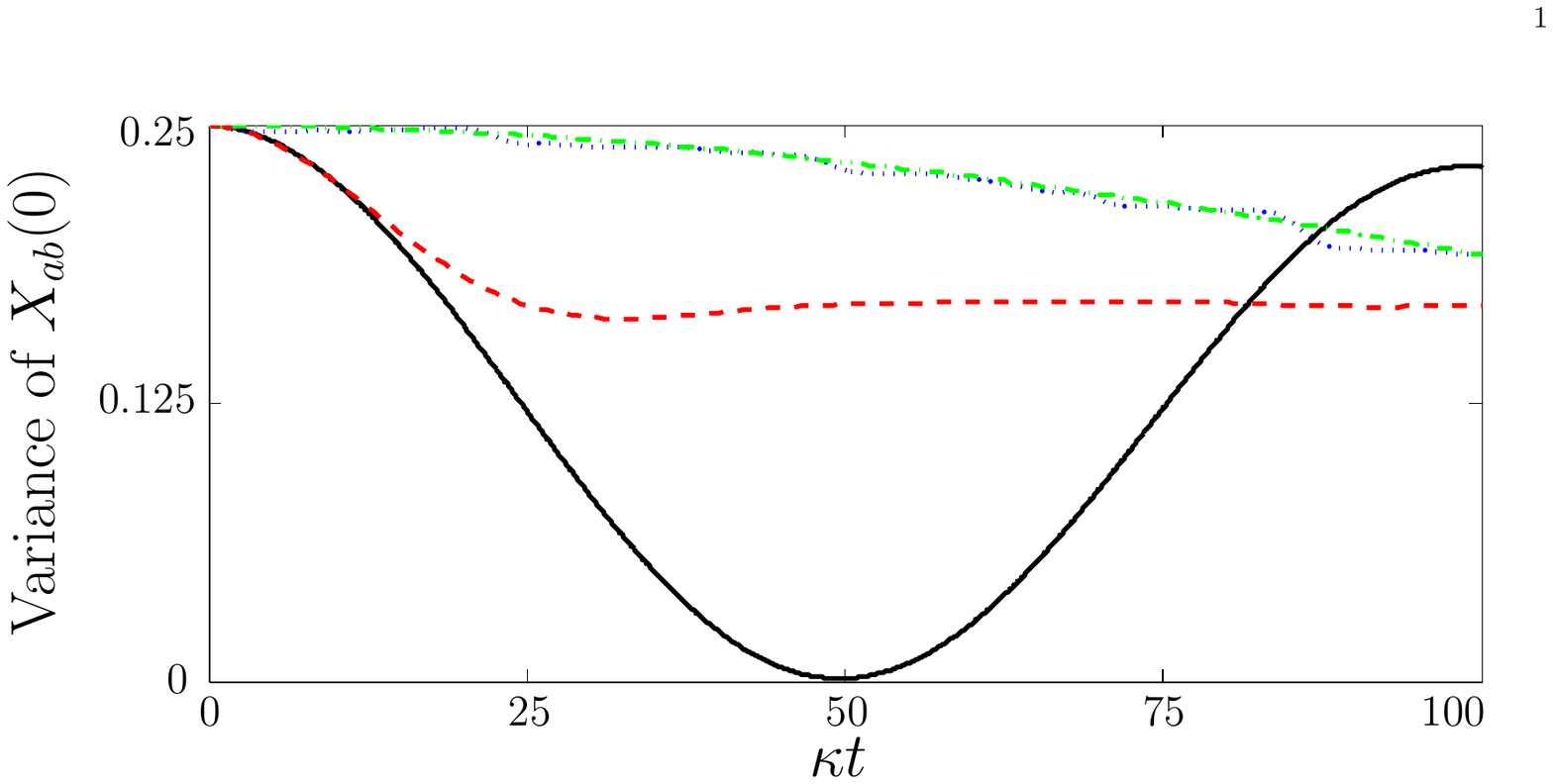}
\includegraphics[width=\linewidth,clip=true,trim=2cm 19.7cm 1.5cm 2cm]{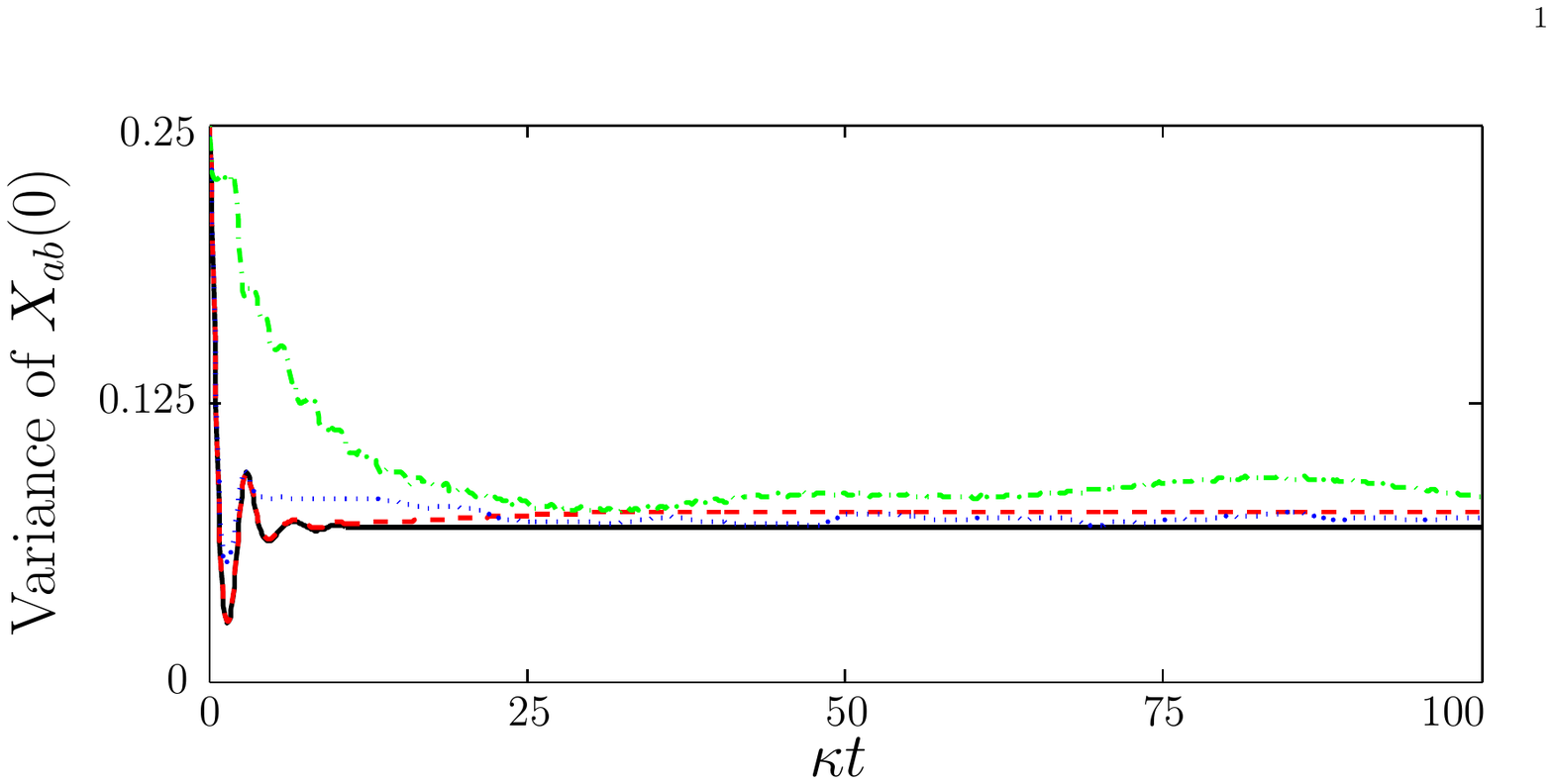}
\caption{The variance of $X_{ab} (0)$ for different broadening width. Defining $g_b = \sqrt{\sum_k |g_{k,b}|^2}$, we have at the top $g_b = 5 \kappa$, $g_a = \sqrt{\sum_k |g_{k,a}|^2} =  0.9g_b$ and $\Delta_c = 75\kappa$ and in the bottom we have the same but $\Delta_c = 0.5\kappa$. The black (solid) line is for a width of $0g_b$. The red (dashed) line is for $0.05\kappa$. The blue (dotted) line is $0.5\kappa$ and the green (dashed dotted) line is for $5\kappa$.} \label{squeezing_inhomo_long}
\end{figure}

\section{Squeezing with inhomogeneity} \label{sec_inhomo}

For solid state spin ensembles, it is not a good approximation to assume perfect degeneracy of the spin excitation energy. Depending on the spin system itself and on the purity of the host material, individual spins have excitation frequencies that may vary within a $MHz$ wide profile or more, and this both detunes the spins from the cavity and causes precession of the spin excitation at different frequencies, leading effectively to a damping of the quadrature correlations.
 
We model the inhomogeneous broadening in the spin ensembles by allowing the energies to be different for each spin, 
\begin{align}
H_0 = \omega_c c\dag c + \frac{1}{2}\sum_{j}^{N_a} \omega_{j,a} \sigma_z^{j,a} + \frac{1}{2} \sum_{j}^{N_b} \omega_{j,b} \sigma_z^{j,b}
\end{align}
and they may also experience different interaction strengths with the quantized field mode
\begin{align}
H_I = \sum_{j}^{N_a} g_{j,a} (\sigma_+^{j,a} c + \sigma_-^{j,a} c\dag)  + \sum_{j}^{N_a} g_{j,b} (\sigma_+^{j,b} c + \sigma_-^{j,b} c\dag).
\end{align}
To use the Holstein-Primakoff formalism we group the spins in frequency intervals with nearly identical frequencies, for which the harmonic oscillator operator representation may be used for the $a$-ensemble,
\begin{align}
a_i = \frac{1}{\sqrt{N_{i,a}}} \sum_{j}^{N_{i,a}}  \sigma_+^{j,a},
\end{align}
and for the $b$-ensemble,
\begin{align}
b_i = \frac{1}{\sqrt{N_{i,b}}} \sum_{j}^{N_{i,b}}  \sigma_-^{j,b}.
\end{align}
With the inverted $a$ spins, we now have the uncoupled and the interaction Hamiltonian
\begin{align}
H_0 = \omega_c c\dag c - \sum_{j}^{\tilde{N}_{a}} \omega_{j} a_j\dag a_j + \sum_{j}^{\tilde{N}_{b}} \omega_j b_j\dag b_j \\
H_I = \sum_{j}^{\tilde{N}_{a}} g_{j,a} (a_j c + a_j\dag c\dag)  + \sum_{j}^{\tilde{N}_{b}} g_{j,b} b_j\dag c + b_j c\dag)
\end{align}
with $\omega_{j,a} = \omega_{j,b} = \omega_j$. 

Passing to a rotating frame, and introducing detunings $\Delta_j$ and $\Delta_c$ with respect to a central spin frequency, we obtain the Heisenberg-Langevin equations
\begin{subequations}
\begin{align}
\frac{d\,a_j}{dt} &= i\Delta_j a_j -i g_{j,a} c\dag \\
\frac{d\,a_j\dag}{dt} &= -i\Delta_j a_j\dag + ig_{j,a} c \\
\frac{d\,b_j}{dt} &= i\Delta_j b_j - i g_{j,b}c\dag \\
\frac{d\,b_j\dag}{dt} &= -i\Delta_j b_j\dag + i  g_{j,b} c \\
\frac{d\,c}{dt} &= -(\kappa + i\Delta_c)c \nonumber \\ & \phantom{= }\,\, -i\sum_j (g_{j,a} a_j\dag + g_{j,b} b_j) - \sqrt{2\kappa}c_{in}(t) \\
\frac{d\,c\dag}{dt} &= -(\kappa - i\Delta_c) \nonumber \\ & \phantom{= }\,\, + i\sum_j (g_{j,a} a_j + g_{j,b} b_j\dag) -\sqrt{2\kappa} c_{in}\dag (t).
\end{align} \label{langevin_inhomo}
\end{subequations}

This leads to deterministic linear equations for the second moments of the quadrature operators, and in Fig. \ref{squeezing_inhomo_long} we show results for the variance of the collective spin mode with different values of the inhomogeneous broadening. In the calculations we have assumed a Gaussian distribution of the spin frequencies. The upper part of the figure shows the case where the cavity is detuned with respect to the spin frequency distribution - the squeezing occurs on a long time scale, and the precession of spins is likely to be the cause of the significant reduction of the squeezing already for moderate broadening. The lower part of the figure shows the almost resonant coupling to the cavity field and a resulting faster squeezing, which is also observed for weak inhomogeneities. In addition, the cavity coupling itself may effectively separate the collectively coupled spin mode from the other modes and thus suppress the effect of inhomogeneities\cite{PhysRevA.53.2711,PhysRevA.83.053852}.

\section{Conclusion} \label{sec_conc}
We have shown that we can create two-mode spin squeezed and entangled states of separate spin ensembles in a microwave cavity and atomic ensembles in an optical cavity from a state where one ensemble is inverted and excitations are transferred between the ensemble via the cavity field.
We have investigated this process with Heisenberg-Langevin equations, which readily provide solutions for the state of the spins and the cavity field, and in the case of a large cavity detuning, we have provided simple expressions for both the long time limit \eqref{variance_long}, and for optimal interaction times where maximum squeezing is obtained \eqref{variance_min}. These results could be investigated in existing systems with trapped atoms and ions in optical cavities, and we have extended the theory to include inhomogeneity in the ensembles, which may be of relevance for electron spin ensembles in solids. Inhomogeneities reduce the squeezing for long times and change the optimum values, but should still be observable, e.g., in experiments with NV centers, and possibly even with species of different particles. We have also shown how the intra cavity squeezing can be transformed into a squeezed pulse of radiation, emitted on demand by the system, and being thus of interest for quantum communication schemes, and possibly also for studies of the interactions between quantum radiation and superconducting qubits\cite{wallraff:2004}. 

\section{Acknowledgement}
The authors acknowledge support from the European Union integrated project AQUTE.

\bibliography{bt}

\end{document}